\documentclass[prl,twocolumn,superscriptaddress,secnumarabic,balancelastpage,amsmath,amssymb]{revtex4-1}
\usepackage{amssymb}
\usepackage{amsmath}
\usepackage[dvipsnames]{xcolor}
\usepackage{chngpage}
\usepackage{lgrind}        
\usepackage{color}         
\usepackage{graphics}      
\usepackage[pdftex]{graphicx}      
\usepackage{longtable}     
\usepackage{epsf}          
\usepackage{bm}            
\usepackage{asymptote}     
\usepackage{thumbpdf}
\usepackage{verbatim}
\usepackage{url}
\usepackage{MnSymbol}
\usepackage{physics}
\usepackage[colorlinks=true]{hyperref} 
\usepackage{enumitem}	

\usepackage{float}
\usepackage{tipa, extraipa}
\usepackage{times}
\usepackage{color}
\usepackage{verbatim}
\usepackage{soul,xcolor}
\setstcolor{red}
\usepackage{colortbl}

\usepackage{tabularx}

\usepackage[normalem]{ulem} 

\newcommand{\outline}[1]{\textit{#1} ---}

\makeatletter
\AtBeginDocument{
\g@addto@macro{\appendix}{\renewcommand{\p@subsection}{\@Alph\c@section}}
}
\makeatother

\begin{document}

\title{Robust Higher-Order Hamiltonian Engineering for\\
Quantum Sensing with Strongly Interacting Systems
}

\affiliation{Department of Physics, Harvard University, Cambridge, Massachusetts 02138, USA}
\affiliation{School of Engineering and Applied Sciences, Harvard University, Cambridge, Massachusetts 02138, USA}
\affiliation{Department of Chemistry and Chemical Biology, Harvard University, Cambridge, Massachusetts 02138, USA}

\author{Hengyun Zhou$^1$, Leigh S. Martin$^1$, Matthew Tyler$^1$, Oksana Makarova$^{1,2}$, Nathaniel Leitao$^1$, Hongkun Park$^{1,3}$, Mikhail D. Lukin$^1$}


\begin{abstract}
Dynamical decoupling techniques constitute an integral part of many quantum sensing platforms, often leading to orders-of-magnitude improvements in coherence time and sensitivity. Most AC sensing sequences involve a periodic echo-like structure, in which the target signal is synchronized with the echo period. We show that for strongly interacting systems, this construction leads to a fundamental sensitivity limit associated with imperfect interaction decoupling. We present a simple physical picture demonstrating the origin of this limitation, and further formalize these considerations in terms of concise higher-order decoupling rules. We then show how these limitations can be surpassed by identifying a novel sequence building block, in which the signal period matches twice the echo period. Using these decoupling rules and the resulting sequence building block, we experimentally demonstrate significant improvements in dynamical decoupling timescales and magnetic field sensitivity, opening the door for new applications in quantum sensing and quantum many-body physics.
\end{abstract}
\maketitle

\outline{Introduction}
Quantum sensing utilizes quantum particles to probe the properties of the surrounding environment~\cite{degen2017quantum}. Recent advances in quantum sensing technologies have led to a host of new applications, including probes of magnetism in condensed matter systems~\cite{casola2018probing,rondin2014magnetometry,du2017control,gross2017real,ku2020imaging}, nuclear magnetic resonance spectroscopy on the nanoscale~\cite{aslam2017nanoscale,glenn2018high}, as well as \textit{in-vivo} temperature sensing~\cite{schirhagl2014nitrogen,kucsko2013nanometre,choi2020probing,fujiwara2020real1}.

\begin{figure}
\begin{center}
\includegraphics[width=\columnwidth]{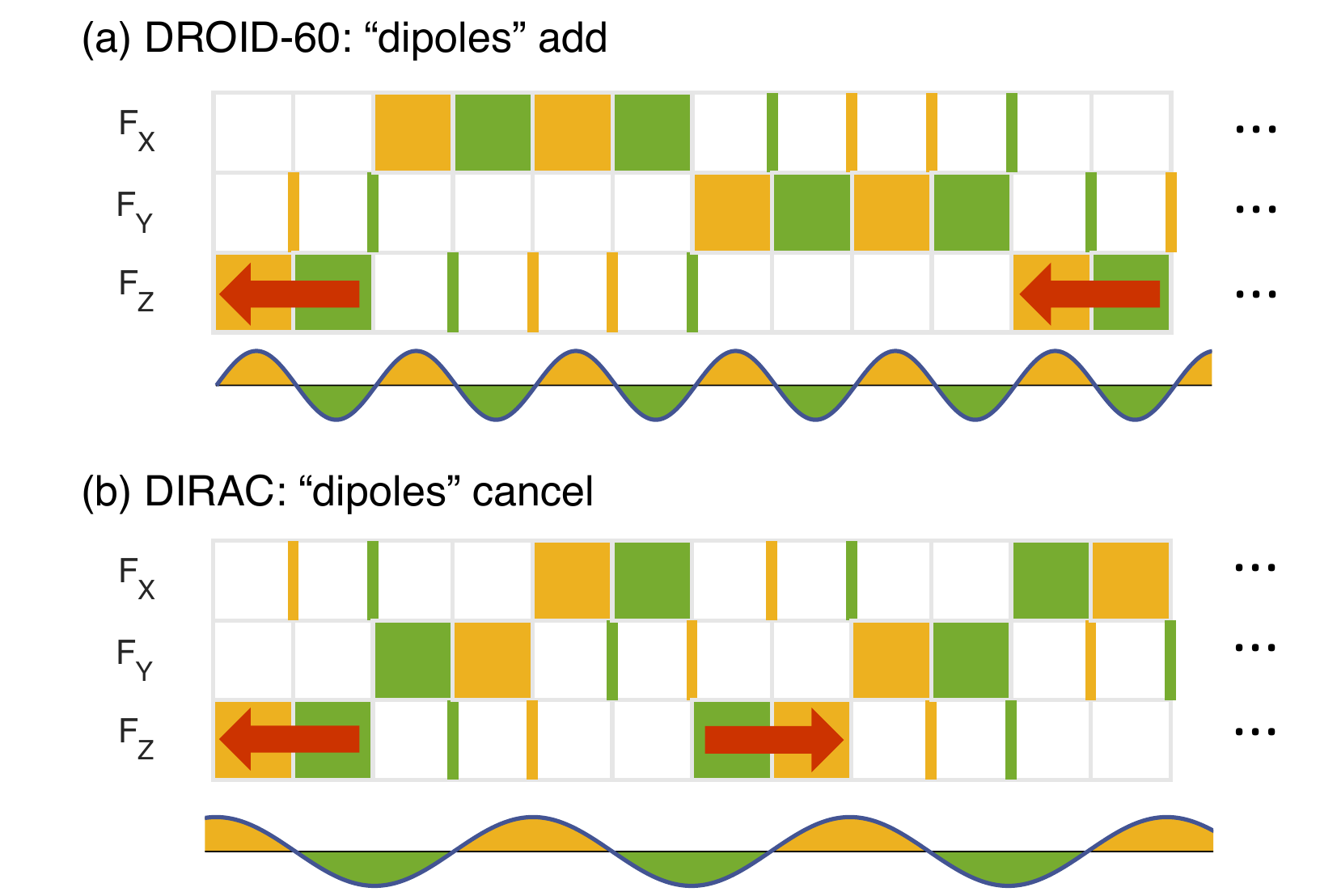}
\caption{{\bf Limits of quantum sensing based on decoupling.} Illustration of the ``dipole" decoupling rule and sensing structure for conventional sensing pulse sequences (DROID-60, a) and our new sensing pulse sequence (DIRAC2, b). $\vec{F}$ represents the orientation of the interaction-picture $\tilde{S}^z$ spin operator, where a yellow (green) block in row $\mu=x,y,z$ signifies $\tilde{S}^z=+S^\mu$ ($-S^\mu$) at the given time point. For XY-8 and DROID-60, maintaining sensitivity to the periodic signal requires dipoles pointing in the same direction (red arrows), such that dipole rules for decoupling cannot be satisfied; this is circumvented by DIRAC, which allows dipole cancellations while maintaining AC field sensitivity.}
\label{hosfig:fig1}
\end{center}
\end{figure}

Key to unlocking new sensing applications is improvements in metrological sensitivity.
A common way to achieve this is to utilize dynamical decoupling sequences~\cite{suter2016colloquium,viola1999dynamical,de2010universal,farfurnik2015optimizing}, such as XY-8 for non-interacting spins~\cite{gullion1990new} or DROID-60 for interacting spin systems~\cite{zhou2020quantum}.
Such sequences typically consist of a train of spin echo pulses synchronized to the target AC field, as illustrated in Fig.~\ref{hosfig:fig1}, which help to isolate the spin system from environmental disorder and certain spin-spin interactions, while maintaining sensitivity to the target signal.
While these techniques are already being actively used across a range of different experimental platforms, their ultimate performance and impact on quantum sensitivity limits are not yet fully understood.

In this Letter, we identify a fundamental limitation associated with the interplay  
between interaction decoupling and sensing in existing pulse sequences, and propose and experimentally demonstrate a class of new pulse sequences that overcomes this limitation.
More specifically, we show that the full synchronization of the sensing signal with a spin echo building block inevitably contradicts the cancellation of higher-order terms in the effective Hamiltonian, which leads to a fundamental limit on the performance of all existing pulse sequences.
To circumvent this limitation, we develop a different paradigm for sensing, in which the target sensing signal period is synchronized with \textit{twice} the spin echo period, allowing one to realize superior sensitivity in experiments.
This is achieved by developing concise decoupling rules for higher-order contributions to the effective Hamiltonian (see Fig.~\ref{hosfig:fig1} for an example), and using them to efficiently screen through large design spaces of pulse sequences.
This results in better pulse sequences for not only quantum sensing but also dynamical decoupling and Hamiltonian engineering, and we find significantly improved performance compared to the best known pulse sequences in the disorder-dominated regime of interacting spins~\cite{zhou2020quantum}.

\outline{Limits on AC quantum sensing based on conventional decoupling} 
The key idea of this work can be understood by considering a disordered, interacting many-body spin system, with the Hamiltonian given by
\begin{align}
H&=H_{signal}(t)+H_{dis}+H_{drive}(t)+H_{int}\nonumber\\
&=B(t)\sum_i S_i^z+\sum_i h_i S_i^z+\sum_i \qty[\Omega_x(t)S_i^x+\Omega_y(t)S_i^y]\nonumber\\&+ \sum_{ij}\left[J^{I}_{ij} S_i^zS_j^z + J^H_{ij} \qty(S_i^xS_j^x+S_i^yS_j^y+S_i^zS_j^z)\right].
\end{align}
This general form encompasses many physical systems, including dipole-dipole interactions, Rydberg atoms, and superexchange-interacting spins. Here, $S_i^{x,y,z}$ are spin-$\frac{1}{2}$ operators, $B(t)$ is the time-dependent sensing target field, $h_i$ is the on-site disorder strength for spin $i$, $\Omega_x(t)$, $\Omega_y(t)$ are time-dependent global Rabi drive strengths for the pulse sequence, $J_{ij}^I$, $J_{ij}^H$ are the Ising and Heisenberg interaction strengths between spins $i$, $j$.
The task of quantum sensing with such an interacting quantum many-body system involves the dual challenge of
\begin{enumerate}
    \item Decoupling strong disorder and interactions as much as possible,
    \item Maintaining maximal sensitivity to the target sensing field under these constraints.
\end{enumerate}

In the interaction picture with respect to the applied drive pulses $\{P_k\}$, due to the secular nature of the Hamiltonian~\cite{choi2020robust}, the Hamiltonian can be expressed purely in terms of a polynomial of the transformed $S^z$ operator (the ``frame") at a given time, i.e.
\begin{align}
    \tilde{H}(t)&=\tilde{H}(\tilde{S}^z(t)),\\
    \tilde{S}^z(t)&=(P_{k-1}\cdots P_1)^\dagger S^z (P_{k-1\cdots}P_1)=\sum_\mu F_{\mu,k}S^\mu,
\end{align}
where the time $t$ is in between the $(k-1)$-th and $k$-th pulse.
For pulse sequences composed of $\pi/2$ and $\pi$ pulses, we denote these ``toggling frame” transformations~\cite{mansfield1971symmetrized} pictorially in Fig.~\ref{hosfig:fig1}, where a yellow/green block in row $\mu=x,y,z$, column $k$ signifies $F_{\mu,k}=+1$ or $-1$ respectively.

Utilizing this representation, we now show that the conventional way of synchronizing the period of the sensing field with the spin echo period necessarily leads to residual first-order terms in the effective Hamiltonian that, in turn, limit the sensing performance. 
These residual terms can be visualized by the ``red dipoles" in Fig.~\ref{hosfig:fig1}(a) for the existing sequence DROID-60, which point from a green block (-1) to a yellow block (+1), and by necessity do not cancel.
Such terms, present in most currently known AC sensing pulse sequences, result in a fundamental limitation for the sensing protocol.

To understand how these terms arise, we make use of the Magnus expansion~\cite{magnus1954on}, which expresses the effective Hamiltonian over one Floquet period $T$ of the pulse sequence as a series summation, with the leading order terms being
\begin{align}
    H^{(0)}&=\frac{1}{T}\int_0^T \tilde{H}(t_1)dt_1,\label{eq:H0}\\
    H^{(1)}&=\frac{-i}{2T}\int_0^T dt_1\int_0^{t_1}dt_2[\tilde{H}(t_1),\tilde{H}(t_2)].\label{eq:H1}
\end{align}
The zeroth-order contribution Eq.~(\ref{eq:H0}) naturally suggests a spin echo structure, since disorder is cancelled when $\sum_k F_{\mu,k}=0$ for each $\mu\in\{x,y,z\}$, such that the positive and negative disorder contributions along each axis balance each other out.
This condition can be interpreted by associating, for each axis $\mu$, a positive charge to a yellow block (+1) and a negative charge to a green block (-1). To cancel disorder,  the sum of charges along each axis must be 0.
Generalizing this analysis to higher-order terms (see Ref.~\cite{tyler2022higher} for more details), we focus on one particular first-order term in Eq.~(\ref{eq:H1}), involving a commutator between on-site disorder and Heisenberg interactions.
Under pulse transformations, the disorder term transforms linearly with $F_{\mu,k}$, while the Heisenberg term is invariant.
Consequently, assuming ideal pulses, we find
\begin{align}
    H_{dis,Heis}^{(1)}&=\frac{-i}{2T}\sum_{\mu,k}\qty(t_k-\frac{T}{2})\tau_k F_{\mu,k}\sum_{i,j,l}\qty[h_i S_i^\mu,J_{jl}^H\vec{S}_j\cdot\vec{S}_l],
\end{align}
where $t_k$ and $\tau_k$ are the center time and duration of the $k$-th free evolution period, respectively.
Focusing on the frame-dependent coefficients, we find that the contribution is 
\begin{align}
H_{dis,Heis,\mu}^{(1)}\propto \sum_k \qty(t_k-\frac{T}{2}) \tau_k F_{\mu,k}.\label{eq:Hdisheis}
\end{align}
Eq.~(\ref{eq:Hdisheis}) describes a sum of dipole moments, since it is the charge $\tau_k F_{\mu,k}$ multiplied by the position $t_k-\frac{T}{2}$.
Thus, we conclude that cancelling the first order cross-term between disorder and Heisenberg interactions requires the sum of dipoles along each axis to cancel.

In light of this interpretation, we can re-examine the pulse sequences previously used for AC field sensing.
In Fig.~\ref{hosfig:fig1}(a), we show the recently-developed sequence for interacting spin systems~\cite{zhou2020quantum}, DROID-60, and how its frames and target AC field are synchronized.
As one can see, the spins flip with the same periodicity as the external magnetic field, and the frames along each axis are always paired up to echo out disorder effects as rapidly as possible.
This immediately implies that for a given axis, the dipoles are always oriented in the same direction in order for the phase accumulation to coherently add, see the red arrows in Fig.~\ref{hosfig:fig1}(a).
Thus, with these pulse sequence structures, maintaining AC field sensitivity always comes at the expense of introducing first order imperfections, which will directly affect the coherence time and subsequently the sensitivity of the sequence.

\outline{Systematic Higher-Order Sequence Design}
To overcome this conflict, we start by systematically incorporating higher-order decoupling conditions (see Ref.~\cite{tyler2022higher} for a full derivation) into sequence design, resulting in dynamical decoupling sequences with much better coherence properties, but are not yet compatible with sensing.
We will then present a novel sequence building block (Fig.~\ref{hosfig:fig1}(b)) that respects the sensing constraints, which allows us to obtain maximal sensitivity to the target sensing field while retaining higher-order decoupling performance.

\begin{figure}
\begin{center}
\includegraphics[width=\columnwidth]{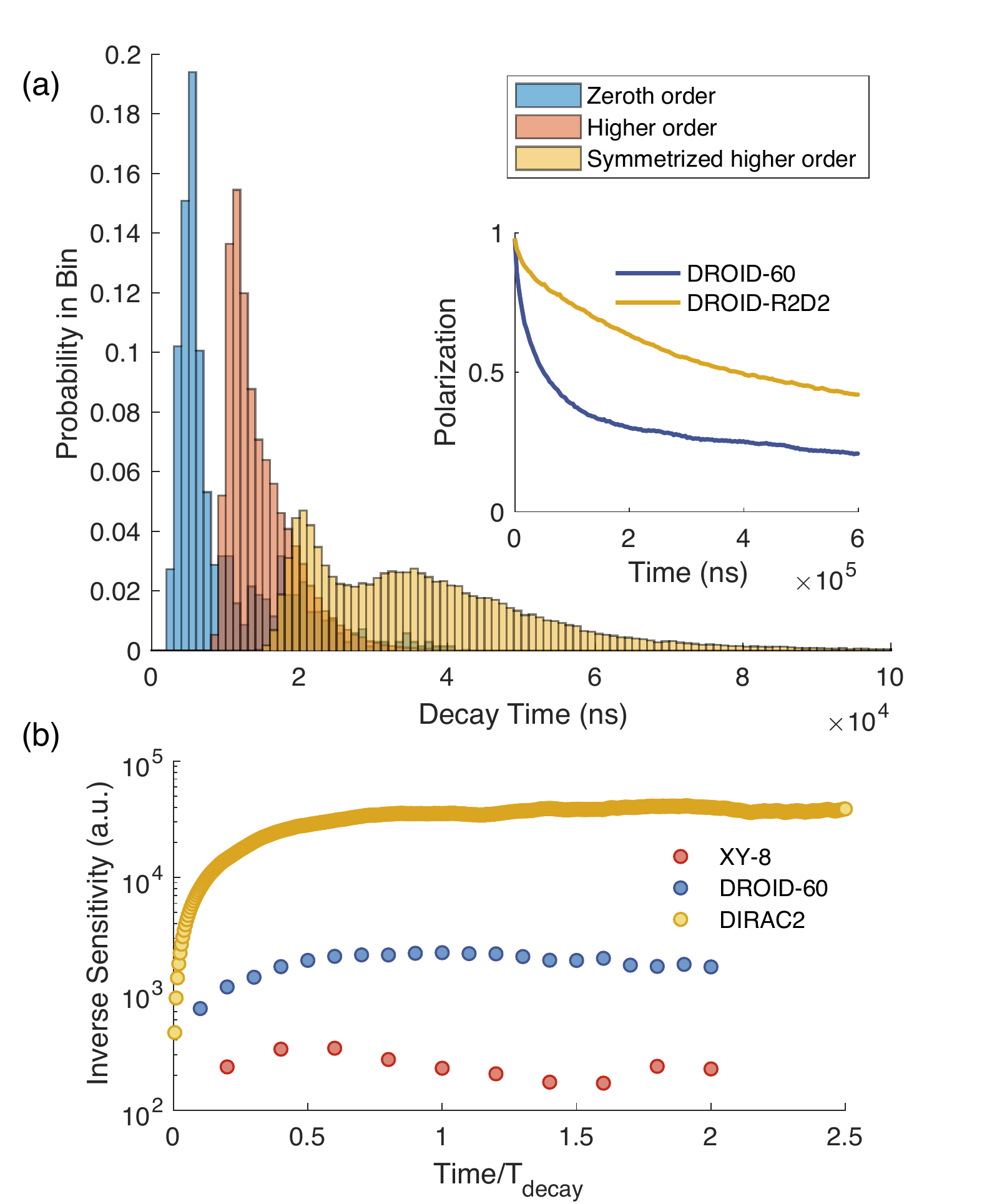}
\caption{{\bf Simulation evaluation and pulse sequence search.} 
(a) Normalized histogram of coherence time, averaged over $\hat{x}$, $\hat{y}$, $\hat{z}$ initial states (averaging performed over decay rates), for different sets of decoupling rules under consideration. Simulations are performed with exact diagonalization of 6 spins. As more rules are sequentially included, the distribution shifts further to the right, eliminating poorly performing pulse sequences.
Inset: Comparison of simulated decoupling performance of the existing sequence DROID-60 and a new sequence DROID-R2D2 that incorporates higher-order terms.
(b) Simulated sensitivities for different pulse sequences, where our new sequence DIRAC2 consistently outperforms all existing sequences. The $x$-axis is normalized by the decay time of the sequence for comparison.}
\label{hosfig:fig2}
\end{center}
\end{figure}

We focus on the regime where the on-site disorder is dominant over spin-spin interactions, as is typically the case for electronic spin ensembles~\cite{zhou2020quantum,zu2021emergent,merkel2021dynamical}.
We numerically simulate the performance of dynamical decoupling pulse sequences designed with different numbers of decoupling rules imposed, using parameters drawn from the experimental system in Ref.~\cite{zhou2020quantum}.
We simulate the decay of a polarized initial state along $\hat{x}$, $\hat{y}$ or $\hat{z}$ after different numbers of repetitions of the full pulse sequence.
The fitted characteristic decay timescales after subtracting out any long-time plateaus (for example, due to residual disorder pinning) are histogrammed in Fig.~\ref{hosfig:fig2}(a) for different sequence design methods, highlighting the progressive improvement in performance as higher-order terms are included.

We start by randomly generating pulse sequences with 24 free evolution frames, where all zeroth-order robust Hamiltonian engineering rules (see Ref.~\cite{choi2020robust}) have been included, but no higher-order rules, as illustrated by the blue bars in Fig.~\ref{hosfig:fig2}(a), resulting in a typical decay timescale on the order of 5~$\mu$s.
Next, we enforce that various higher-order terms are zero (see Ref.~\cite{tyler2022higher}), with the most crucial ones being first-order cross-terms between disorder and Heisenberg interactions (the ``dipole rule" described above), as well as second-order terms originating purely from disorder.
This significantly reduces the sequence search space, allowing us to exhaustively search through pulse sequences up to length 24.
This extends the decay time out to the orange bars.
Crucially, it removes the sequences that had relatively short decay times, revealing the longer lived ones.
Finally, we apply a further layer of symmetrization to the pulse sequence, where the frame ordering is reversed and sign flipped in the second half of the sequence.
This further improves the decay times, as seen in the yellow distribution of Fig.~\ref{hosfig:fig2}(a), with a long tail extending to the right.
The longer timescales also imply that certain higher-order imperfections that may be dominant when performing general Hamiltonian engineering are also systematically removed.

In the inset of Fig.~\ref{hosfig:fig2}(a), we show a direct comparison between the previous best sequence in this parameter regime, DROID-60~\cite{zhou2020quantum}, designed with zeroth-order rules and the same symmetrization methods, and a new sequence DROID-R2D2 (Disorder RObust Interaction Decoupling - Robust To Disorder 2nd order) that accounts for higher-order rules.
As one can see, the best symmetrized higher-order pulse sequences show almost an order of magnitude improvement in decay time compared to prior sequences that only include zeroth-order terms (for the simulation parameters mentioned above, DROID-R2D2 has $1/e$ decay time 390 $\mu$s, compared to 64 $\mu$s for DROID-60).
This highlights the power of systematically including higher-order rules in the sequence design.

\outline{Surpassing the AC Sensing Limit with Higher-Order Sequence Design}
Although incorporating higher-order decoupling conditions improves coherence times for dynamical decoupling, it does not fully overcome the conflict between sensing and interaction decoupling.
This is because as described above, when the spins flip with the same periodicity as the magnetic field, maintaining AC field sensitivity always results in residual first-order imperfections in the form of ``dipole-rule" violations in Eq.~(\ref{eq:Hdisheis}).

To address this challenge, we devise a new strategy for the design of AC sensing pulse sequences that overcomes this conflict between sensing and decoupling.
Instead of requiring the frame flips to be commensurate with the AC signal, the key idea is to design the frame flips to be at twice the rate of the AC signal.
By moving between frames on different axes, it is possible to continue to coherently accumulate phase for interaction-decoupling AC sensing sequences in this case.

Fig.~\ref{hosfig:fig1}(b) illustrates a representative sequence designed in this way, which we name DIRAC2 (DIsorder Robust AC sensing of period 2).
In this sequence, the yellow and green blocks for the frame matrix are lined up with positive and negative values of the target AC signal, respectively, indicating coherent phase addition and maximal sensitivity.
In addition, the dipoles along each axis cancel each other out, indicating that first-order disorder-Heisenberg interaction cross-terms mentioned in Eq.~(\ref{eq:Hdisheis}) are fully suppressed.
Moreover, the faster flipping rate of the frames relative to the AC signal has the added benefit that for the same target signal, this pulse sequence is more effective in decoupling time-varying noise, which can lead to further performance improvements.
Finally, this sequence also incorporates a number of other higher-order considerations and symmetrizations described in the previous section to further boost performance.
Indeed, in Fig.~\ref{hosfig:fig2}(b), we simulate the performance of different decoupling sequences with the same parameters as the preceding section, where we find that DIRAC2 can outperform the best known decoupling sequences for interacting spin ensembles (DROID-60) by an order of magnitude.

\begin{figure}
\begin{center}
\includegraphics[width=\columnwidth]{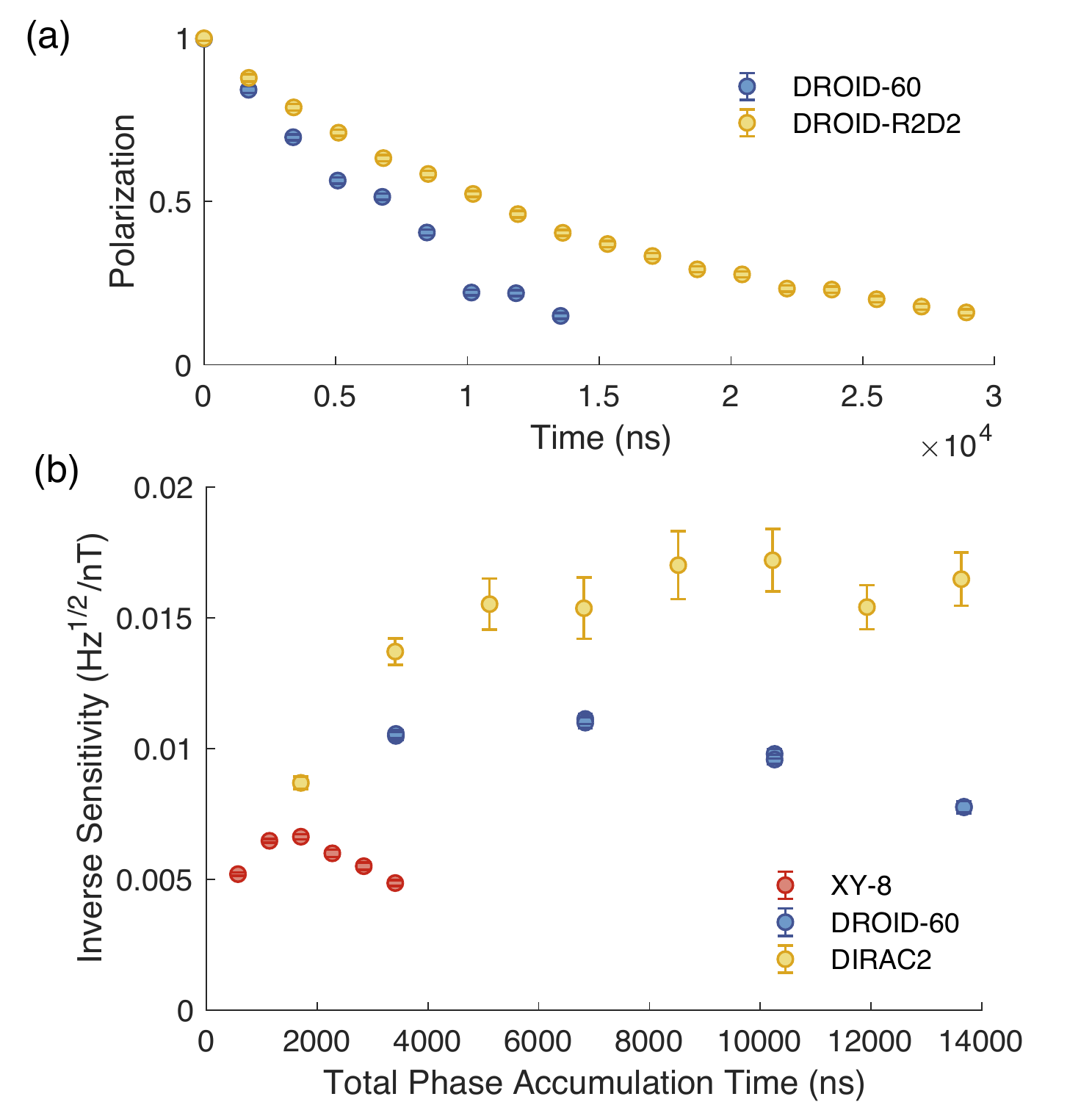}
\caption{{\bf Experimental improvements for dynamical decoupling and sensing with higher-order pulse sequence design.}
(a) Experimental coherence decay curves under dynamical decoupling with different sequences, where the new sequence DROID-R2D2 significantly outperforms the existing sequence DROID-60.
(b) Experimental AC magnetic field sensitivities, where our new sequence DIRAC2 is significantly better than DROID-60 and XY-8, improving quantum sensing with strongly-interacting spin ensembles.}
\label{hosfig:fig3}
\end{center}
\end{figure}

\outline{Experimental Performance of Higher-Order Sequences}
To verify the performance of our new methods, we experimentally implement these decoupling and sensing sequences in a high-density ensemble of nitrogen-vacancy (NV) centers in diamond, see Ref.~\cite{kucsko2018critical,zhou2020quantum} for more details of the experimental system.
First, we compare the performance of our new dynamical decoupling sequence DROID-R2D2 against the previous best sequence DROID-60 in Fig.~\ref{hosfig:fig3}(a), where we prepare an $\hat{x}$ initial state, decouple with either the DROID-60 or DROID-R2D2 sequence for a period of time, and measure the final polarization along the $\hat{x}$ axis.
The total measurement window is held constant to normalize out charge dynamics and $T_1$ decay effects, and focus on decoupling performance.
The higher-order sequence DROID-R2D2 that we design here considerably outperforms the best known sequence DROID-60 (15 $\mu$s vs. 9 $\mu$s), although technical imperfections still limit the achievable coherence times to be shorter than simulations for both sequences.

Having shown that higher-order sequence design improves dynamical decoupling, let us now compare the performance of different sensing sequences.
For the comparison, we choose a target signal frequency of $(2\pi)\times 7$ MHz, a $\pi$ pulse time of 10 ns, and a free evolution time $\tau=61$ ns for XY-8, DROID-60, $\tau=25$ ns for DIRAC2, in order to synchronize the sequence and sensing signal.
In Fig.~\ref{hosfig:fig3}(b), we show the inverse sensitivity (the higher the better) for each of the sequences, as a function of the total phase accumulation time.
The sensitivities under optimal measurement conditions in this sample, \textit{i.e.} the highest point on each curve, were $\eta_{\textrm{XY-8}}=151\pm 2$  $\textrm{nT}/\sqrt{\textrm{Hz}}$ for the XY-8 sequence, $\eta_{\textrm{DROID}}=90\pm 2$  $\textrm{nT}/\sqrt{\textrm{Hz}}$ for the DROID-60 sequence, and $\eta_{\textrm{DIRAC}}=58\pm 4$  $\textrm{nT}/\sqrt{\textrm{Hz}}$ for the DIRAC2 sequence.
Thus, while DROID-60 outperforms the non-interacting sensing sequence XY-8, DIRAC2 achieves yet another significant improvement over DROID-60, achieving close to a factor of 3 improvement in sensitivity over the conventional sensing sequence XY-8.
This is because the combination of suppressed higher-order terms and faster decoupling results in longer coherence times, while maintaining a similar rotation rate under the target field.

\outline{Discussions and Outlook}
In this work, we identified a key limitation of existing pulse sequences for dynamical-decoupling-based quantum sensing, and devised a novel higher-order Hamiltonian engineering method to overcome these limitations.
We implemented these sequences experimentally, resulting in significant improvements in AC magnetic field sensitivities using these techniques.

These gains in sensitivity can be immediately translated to nanoscale NMR experiments, as recently demonstrated with DROID-60 in Ref.~\cite{arunkumar2022quantum}.
In addition to extending coherence times for sensing sequences, these techniques are also important in extending coherence times for dynamical decoupling and removing systematic artifacts when engineering desired target many-body Hamiltonians~\cite{martin2022controlling,tyler2022higher}.

The analytical insights developed here provide simple geometric intuitions for various higher-order decoupling rules, which significantly simplify the design and optimization process. We expect that our techniques can be readily extended to even higher-order contributions.
Moreover, it will also be interesting to explore the application of these ideas to spin systems in other parameter regimes, such as Rydberg atoms~\cite{bluvstein2021controlling,geier2021floquet}, nuclear magnetic resonance~\cite{cory1990time,peng2021deep}, and trapped ions~\cite{zhang2017observation,kokail2019self}, or higher spin systems~\cite{zhou2022robust,leitao2022qudit}.
Finally, our results provide an important tool for the reliable engineering of many-body Hamiltonians, free of higher-order artifacts, opening the door to exploration of exotic driven phases of matter and creation of entangled quantum states for quantum metrology~\cite{davis2016approaching,hosten2016quantum,cappellaro2009quantum,goldstein2011environment}.

\outline{Acknowledgements}
We thank J.~Choi, H.~Gao, N.~Maskara for helpful discussions. This work was supported in part by CUA, ARO MURI, DARPA DRINQS, Moore Foundation GBMF-4306, NSF PHY-1506284.

\bibliography{main}

\end{document}